\documentclass[%
 reprint,
 amsmath,amssymb,
 aps,
showpacs,
floatfix,
natbib,
prl,
]{revtex4-1}
\usepackage{graphicx}% Include figure files
\usepackage{dcolumn}% Align table columns on decimal point
\usepackage{bm}% bold math

\begin{document}

\title{Plastic yielding and deformation bursts\\ in the presence of disorder from coherent precipitates}

\author{Henri Salmenjoki$^1$, Arttu Lehtinen$^1$, Lasse Laurson$^{2}$ and  Mikko J.\  Alava$^{1,3}$}
\affiliation{$^1$Aalto University, Department of Applied Physics, PO Box 11000, 00076 Aalto, Finland\\
$^2$Computational Physics Laboratory, Tampere University, P.O. Box 692, FI-33101 Tampere, Finland\\
$^3$NOMATEN Centre of Excellence, National Centre for Nuclear Research, A. Soltana 7, 05-400  Otwock-Swierk, Poland}

\begin{abstract}
Alloying metals with other elements is often done to improve the material strength or hardness. A key microscopic mechanism is precipitation hardening, where precipitates impede dislocation motion, but the role of such obstacles in determining the nature of collective dislocation dynamics remains to be understood.
Here, three-dimensional discrete dislocation dynamics simulations 
of FCC single crystals are performed with fully coherent spherical precipitates from zero precipitate density upto  
$\rho_p = 10^{21}\,\text{m}^{-3}$ and at various dislocation-precipitate interaction strengths.
When the dislocation-precipitate interactions do not play a major role 
the yielding is qualitatively as for pure crystals, i.e., dominated by "dislocation jamming", resulting in glassy dislocation dynamics exhibiting critical features at any stress value. We demonstrate that increasing the precipitate density and/or the dislocation-precipitate interaction strength 
creates a true yield or dislocation assembly depinning transition, 
with a critical yield stress. This is clearly visible in the statistics 
of dislocation avalanches observed when quasistatically ramping up the
external stress, and is also manifested in the response of the system to constant applied stresses. The scaling of the yielding with precipitates 
is discussed in terms of the Bacon-Kocks-Scattergood relation.
\end{abstract}

\maketitle

\section{Introduction}
Crystalline materials accumulate plastic deformation via the motion of
dislocations, the line-like topological defects of the crystal
lattice. Hence, controlling the stress-driven dynamics of dislocations 
is the key to be able to tune the mechanical properties of crystals. 
Alloys formed by mixing other elements to pure metals often exhibit increased 
strength or hardness. An important microscopic mechanism routinely exploited in metallurgy is precipitation hardening, where small particles formed 
by the alloying element obstruct dislocation motion \cite{ardell1985precipitation,beyerlein2019alloy}. 

The emergence of experimental techniques
such as compression of micron-scale samples with nanoindentors 
\cite{uchic2004sample,dimiduk2006scale,stinville2019dislocation,dehm2018overview} and high-resolution acoustic emission (AE) measurements of bulk samples \cite{miguel2001intermittent} has revealed a novel paradigm:
dislocation plasticity is a spatio-temporally 
fluctuating and intermittent process \cite{zaiser2006scale}.
On micron scales, discrete strain bursts with a broad size distribution can be seen directly in the stress-strain curve \cite{ng2008stochastic,
papanikolaou2017avalanches,salmenjoki2018machine,zhang2017taming}. 
Macroscopic samples tend to exhibit a smooth stress-strain curve, 
but AE measurements show acoustic energy bursts 
spanning several orders of magnitude in energy \cite{miguel2001intermittent,weiss2015mild}.
The observed strain bursts and AE events originate 
from the stress-driven 
cooperative rearrangements within the crystal, 
known as dislocation avalanches \cite{csikor2007dislocation,zaiser2006scale,dimiduk2006scale,sparks2018nontrivial}. In general, stress-strain curves and hence the global plastic response of crystals consist of a sequence of such avalanches, separated by regions of nearly linear, quasireversible deformation \cite{szabo2015plastic}.

Recent discrete dislocation dynamics (DDD) studies
of pure %or weakly disordered 
crystals (i.e., crystals with no other defects or impurities in addition to dislocations) have revealed a dislocation jamming 
\cite{miguel2002dislocation} dominated regime characterized by 
``extended criticality'' \cite{ispanovity2014avalanches,lehtinen2016glassy,ovaska2017excitation}. Such "glassy" material response was found first in 2D \cite{ispanovity2014avalanches} and then by 3D DDD
simulations \cite{lehtinen2016glassy}. The main quantity of interest here is the size distribution of plastic slip avalanches, which is often found to be well-described by a power-law with a cut-off as 
\begin{equation}
\label{eq:1}
P(s| \sigma, L) \propto s^{-\tau_s} f\left(s/s_0(\sigma,L)\right),
\end{equation}
where $f(x)$ is a cut-off scaling function, typically an exponential, $f(x) \propto \textrm{e}^{-x}$, $s$ is a measure of the avalanche size, $s_0$ is the cut-off avalanche size, $\sigma$ the external stress, and $L$ the system size.
The main signature of extended criticality is power-law like behavior of $P(s| \sigma, L)$ as described in Eq. (\ref{eq:1}), with a particular scaling of the cut-off avalanche size: $s_0$ exhibits an exponential dependence on $\sigma$ and (when considering an "extensive" definition of $s$ \cite{ispanovity2014avalanches}) a power-law divergence with $L$. Thus, there is no special, "critical" value $\sigma=\sigma_c$ where $s_0$ would diverge, contrary to what one would expect in the context of systems exhibiting a (non-equilibrium) phase transition between "jammed" and "moving" phases. Instead, the divergence of $s_0$ with $L$ implies that the system is critical at any stress, hence the notion of extended criticality \cite{ovaska2017excitation,lehtinen2016glassy,ispanovity2014avalanches}. 
It is also known that analogous phenomenology is found for creep (constant stress) simulations. The strain rate fluctuates quite some in a typical (small) sample, but regardless of $\sigma$ follows an Andrade-like scaling
$\varepsilon_t \propto t^{-\theta}$ \cite{miguel2002dislocation} for a considerable range of time which increases with the system size \cite{ispanovity2011criticality}. In contrast, for typical non-equilibrium phase
transitions the order parameter (strain rate for yielding) would decay exponentially in time unless one sets the control
parameter, the stress, close to the critical point value. The variation of the avalanche size distribution with applied stress or level of plastic deformation is also visible in the envelope (the averaged shape) of the intermittent stress-strain curves \cite{szabo2015plastic}.

The extended criticality scenario described above is expected to apply also to crystals with weak but non-zero pinning due to precipitates, such that the dislocation-dislocation interactions dominate the dislocation dynamics. Stronger precipitate pinning as studied here would in principle be expected to change this picture. 
If the precipitate-dislocation becomes the dominating interaction over the dislocation-dislocation (jamming), one may expect in analogy to 2D DDD simulations \cite{ovaska2015quenched} that a well-defined critical point of a critical phase transition would ensue, with collective dislocation dynamics only in the proximity of the critical point. 
Hence, the main feature of avalanches in plasticity in that case would be power-law scaling of $P(s|\sigma,L)$ with a cut-off $s_0$ diverging at a critical stress $\sigma_c$ only. Similarly, the response of the system to constant applied stresses might be expected to show signatures of a depinning transition: At the critical point $\sigma=\sigma_c$, power-law relaxation in time of the strain rate is expected, $\varepsilon_t \propto t^{-\theta}$, again in analogy to recent 2D DDD results \cite{ovaska2015quenched}.  
For weak enough stresses the strain rate relaxation
should be exponential in time, and for $\sigma$ slightly above the critical value one expects a cross-over to a state of
continuous flow after a relaxation transient. We also point out that the 2D DDD results of Ref. \cite{ovaska2015quenched} suggest that very strong disorder-induced pinning may completely quench collective dislocation dynamics. This finding may be related to the concept of mild vs wild fluctuations in crystal plasticity when the microstructure is varied among others by controlling the disorder \cite{weiss2015mild,zhang2017taming}.

\begin{figure*}[hbt!]
  \centering
    \includegraphics[width=0.8\textwidth]{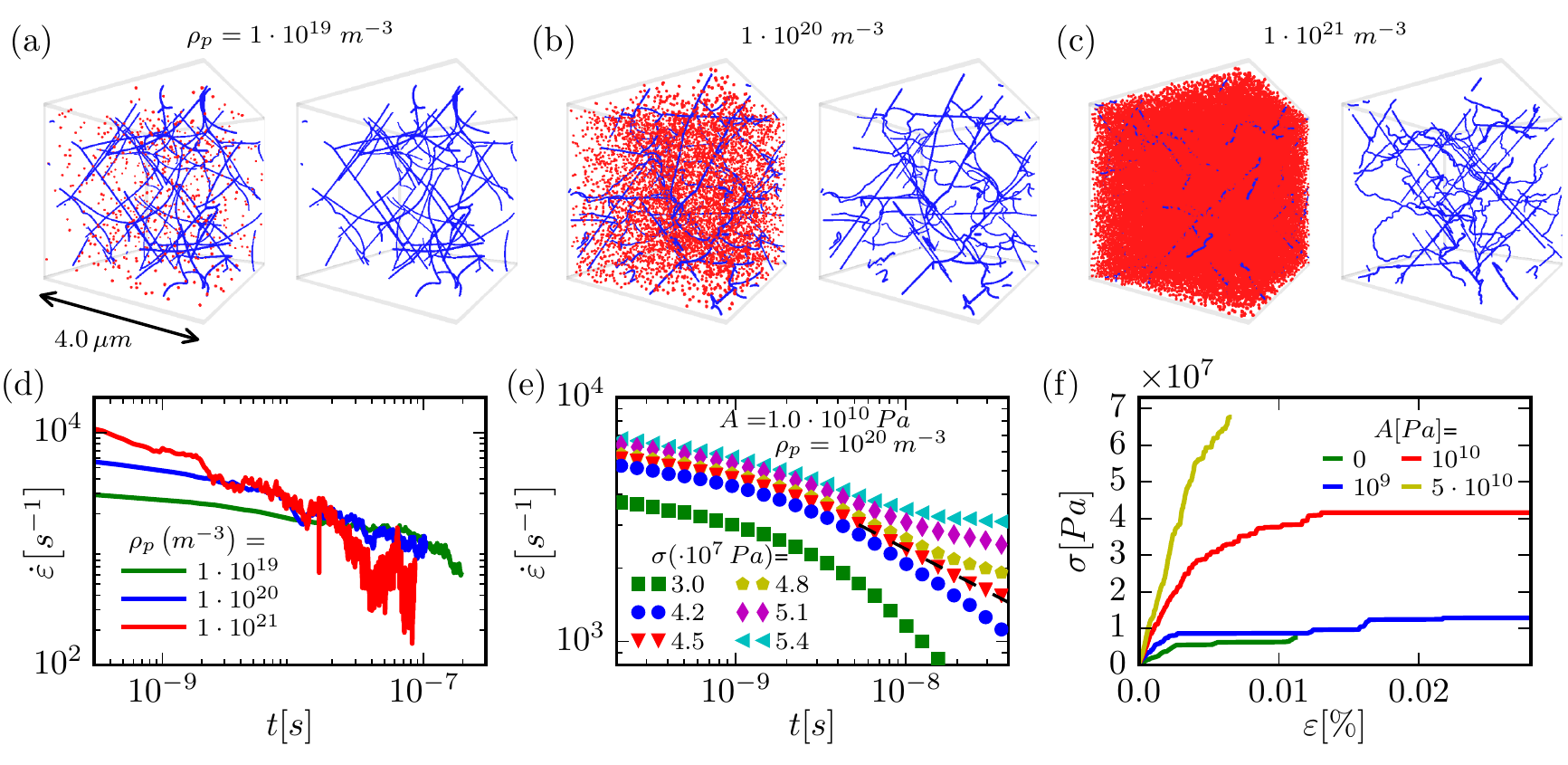}
  \caption{(a)-(c) Snapshots of dislocation
  systems at $t=7 \cdot 10^{-8}\, \text{s}$ from the constant stress 
    simulations with $\sigma$ close to the yield stress $\sigma_{c'}(\rho_p)$ and precipitate 
    parameters $A=1.0 \cdot 10^{10}\,\text{Pa}$ and 
    (a) $\rho_p=10^{19}\,\text{m}^{-3}$, 
    (b) $\rho_p=10^{20}\,\text{m}^{-3}$, and 
    (c) $\rho_p=10^{21}\,\text{m}^{-3}$. 
   (d)  strain rate $\dot{\varepsilon}$ vs $t$ for the illustrated systems. 
   (e)  $\sigma_{c'}$ are determined by $\sigma$ that produce 
    power-law relaxation $\dot{\varepsilon}\sim t^{-\theta}$, here $\sigma_{c'} = (4.5 \pm 0.3) \cdot 10^{7}\, \text{Pa}$. (f) Example stress-strain curves of quasi-statically driven systems with fixed $\rho_p=10^{20}\ \text{m}^{-3}$ and varying interaction strength $A$.  }
  \label{fig1}
\end{figure*}

These considerations then lead us to the question of how one may by controlling the precipitate content of a 3D crystal tune its mechanical properties \cite{zhang2017taming,csikor2007dislocation,pan2019rotatable} and how that relates
to the statistical physics aspects - relaxation, burst size distributions - for a given precipitate microstructure.
Previous studies \cite{monnet2011orowan,xu2016edge,keyhani2018dislocation,santos2018discrete} have addressed the problem of {\it individual} dislocations interacting with precipitates. Here, we focus on understanding the {\it collective} dynamics of many dislocations interacting with a precipitate-induced pinning field. To this end,
we study how small, randomly distributed spherical obstacles of different dislocation-precipitate interaction strengths $A$ and densities $\rho_p$ affect the collective dislocation dynamics in FCC aluminum single crystals. This is done by extensive 
3D DDD simulations utilizing our recently developed methodology \cite{lehtinen2016multiscale,lehtinen2018effects} to include coherent precipitates with short-range elastic interactions with the dislocations \cite{lehtinen2016multiscale,lehtinen2018effects,arsenlis2007enabling}. We consider dislocation networks under both quasistatically increasing stress and constant loading as illustrated in Fig.\ \ref{fig1}. With increasing pinning strength, in analogy to previous results on simplified 2D DDD models \cite{ispanovity2014avalanches,ovaska2015quenched}, our 3D DDD results show a cross-over between two distinct
regimes of material response, arising from the competition of 
dislocation-dislocation and dislocation-precipitate interactions. 

In the rest of this article, after explaining the methods and the choice of parameter space we turn to the results. First, we discuss the pinning strength dependent properties on the relaxation dynamics. We expect that the two phases influence the constant stress response in a fundamental way, and
demonstrate this. We also analyze the dependence of the ensuing yield stress of the crystal on the precipitate density in terms of the Bacon-Kocks-Scattergood relation \cite{bacon1973effect} and how that relates to the phases, jamming against precipitate-induced depinning. After that we move to quasistatic loading, by a stress ramp. We show how the
precipitates influence stress-strain curves, the engineering yield stress, and the statistical properties of the strain bursts occurring during such loading. We also demonstrate that the yield stress values ("critical point value") that we derive from the two approaches agree.

\section{Methods}
Our simulations are performed using a modified version of the ParaDiS 3D DDD code \cite{arsenlis2007enabling, lehtinen2016multiscale,lehtinen2016glassy} incorporating in addition to the dislocations also a description of spherical obstacles or precipitates. The dislocation-precipitate interaction is parametrized with $r_p = 28.6\,\text{nm} $, the radius of the precipitate, and $A$, the factor scaling the precipitate strength in the radial force caused by Gaussian interaction potential
\begin{equation}
\label{eq:fdp}
F(r) = - \nabla U(r) = \frac{2 A b^3 r e^{-\frac{r^2}{r_p^2}}}{r_p^2},
\end{equation}
where $b$ is the length of the Burgers vector.
Notice that the isotropic, short-range interaction force of Eq. (\ref{eq:fdp}) mimics fully coherent precipitates which do not induce long-range stress fields within the embedding crystal.
$A$ and the density of precipitates $\rho_p$ act as control parameters; the latter sets a lengthscale that breaks the similitude principle of pure dislocation systems \cite{zaiser2014scaling}. For a given embedding crystal containing specific types of precipitate particles, the $A$ parameter (as well as $r_p$) can be estimated by comparing DDD and molecular dynamics (MD) simulations \cite{lehtinen2016multiscale}. In practice this is done by measuring the stress required to drive a dislocation line through the precipitate in MD simulations, after which $A$ is adjusted in DDD to reproduce the result. A high value means simply an impenetrable precipitate \cite{santos2020AlCu}. It is also possible to use experimental input, e.g., high-resolution transmission electron microscopy measurements of the elastic fields of the precipitate, which can subsequently be used in DDD simulations \cite{ringdalen2017dislocation}. In general, the Gaussian potential allows tuning of the defect strength from weak, shearable obstacles to strong impenetrable obstacles which the dislocations have to bypass via the Orowan mechanism. In the former case, the precipitates are here for simplicity taken to remain intact even after a dislocation line has moved through them. A final note is that real precipitate-host systems have a size distribution for the precipitates, and not all of the precipitating species necessarily ends in the precipitates thus changing the dislocation mobility. These effects we neglect, though certainly it would be an interesting exercise to add such detail to the DDD simulations from experiment and multiscale simulations.

The material parameters employed here mimic FCC Al with precipitates \cite{lehtinen2016glassy}.
The system was implemented with periodic boundaries and size $L=4\,\mu \text{m}$ which is sufficiently large to produce behaviour independent of the system size (Supplementary Fig.\ 1 \cite{SM}). One may notice that this size scale is close to typical micropillar experiments \cite{sparks2018nontrivial,zhang2017taming,pan2019rotatable,ng2008stochastic}, but here the use of periodic boundaries implies that we are simulating a part of a bulk system.
24 initial dislocations in the slip system $\frac{1}{2}\langle 110 \rangle \{ 111\}$ imply an initial dislocation density $\rho_0 \approx 2.0 \cdot 10^{12}\,\text{m}^{-2}$.
The edge and screw dislocation mobilities, $M_{\text{edge}}$ and $M_{\text{screw}}$, are chosen as $10^{4}\, (\text{Pa}\,\text{s})^{-1}$.
The precipitate strength $A$ ranges from (pure dislocation systems with $0$ to) $1.0 \cdot 10^{9} 
\,\text{Pa}$ (weak) and $1.0 \cdot 10^{10}\,\text{Pa}$ (intermediate) to $5.0 \cdot 10^{10}\,\text{Pa}$ 
(strong precipitates). Only the strongest precipitates are able to cause dislocations to form Orowan loops \cite{lehtinen2016multiscale}.
$\rho_p$ is varied from 
$10^{18}\,\text{m}^{-3}$ to $10^{21}\,\text{m}^{-3}$ 
(i.e. at most $N_p = 65508$ precipitates in the system).
The volume fraction ($\leq 10 \%$) corresponds to relevant experiments \cite{zhang2017taming}.
Other relevant simulation parameters are collected to Supplementary Table I \cite{SM}.

Constant stress DDD simulations allow to observe the power-law relaxation of $\dot{\varepsilon}$ (or lack thereof), and to locate the yield stress by searching for a $\sigma$-value resulting in pure power-law behavior of $\dot{\varepsilon}$. After initialization (relaxation and a second step of relaxation after adding the precipitates), a constant stress $\sigma$ is applied in the $\left[100\right]$ direction. To find the approximate value of the critical stress $\sigma_{c'}$, systems with different $\rho_p$ are loaded with 4 to 8 different values of $\sigma$ with 19 simulations for every $(\rho_p,\sigma)$ combination. 

To obtain the avalanche statistics, a quasi-static stress ramp driving protocol is used \cite{lehtinen2016glassy}. Here, this amounts to imposing a stress rate $\dot{\sigma} = 1.0 \cdot 10^{14}\,\text{Pa/s}$ in the $\left[100\right]$ direction in between avalanches while keeping $\sigma$ constant during avalanches. To define avalanches, we consider the "activity signal" defined by the extensive dislocation 
velocity, $V(t) = \sum_i l_i v_{\bot,i}$, where $l_i$ is the segment length 
and $v_{\bot,i}$ is the velocity perpendicular to the segment's line 
direction. Then, an avalanche is defined as a continuous sequence of $V(t)$-values exceeding a threshold value $V_{\text{thres}}= 5 \cdot 10^{-6}\, \text{m}^2/\text{s}$. The avalanche size $s$ is then defined as $s = \int_0^T (V(t)-V_{\text{thres}})\mathrm{d}t$, 
where $T$ denotes the duration of the avalanche. Here we fix the precipitate density to $\rho_p= 10^{20}\, \text{m}^{-3}$ and vary the precipitate strength $A$, and collect statistics from 100 different initial dislocation-precipitate configurations.

\section{Plastic flow under constant stress}
The creep-like flow under constant stress is studied next, for 
a relaxed configuration of initially straight dislocations (initial 
dislocation density $\rho_0 \approx 2.0\cdot 10^{12}$ m$^{-2}$).
Example snapshots of deforming systems 
for different densities $\rho_p$ of precipitates of intermediate strength
are depicted in Fig. \ref{fig1}(a)-(c). A clear effect of increasing $\rho_p$ on 
the structure of the dislocation network is evident: In Fig. \ref{fig1}(a), 
displaying a system with $\rho_p = 10^{19}$ m$^{-3}$, only a few of the 
dislocation segments are pinned by the precipitates, but increasing 
$\rho_p$ to $10^{20}$ m$^{-3}$ [Fig. \ref{fig1}(b)] and even to 
$10^{21}$ m$^{-3}$ [Fig. \ref{fig1}(c)] results in a noticeable
increase of pinning and hence roughening of the dislocation
lines due to precipitates. In what follows, these morphological changes due to precipitates are related to the collective dynamics and plasticity.

 \begin{figure}[t!]
  \centering
    \includegraphics[width=0.38\textwidth]{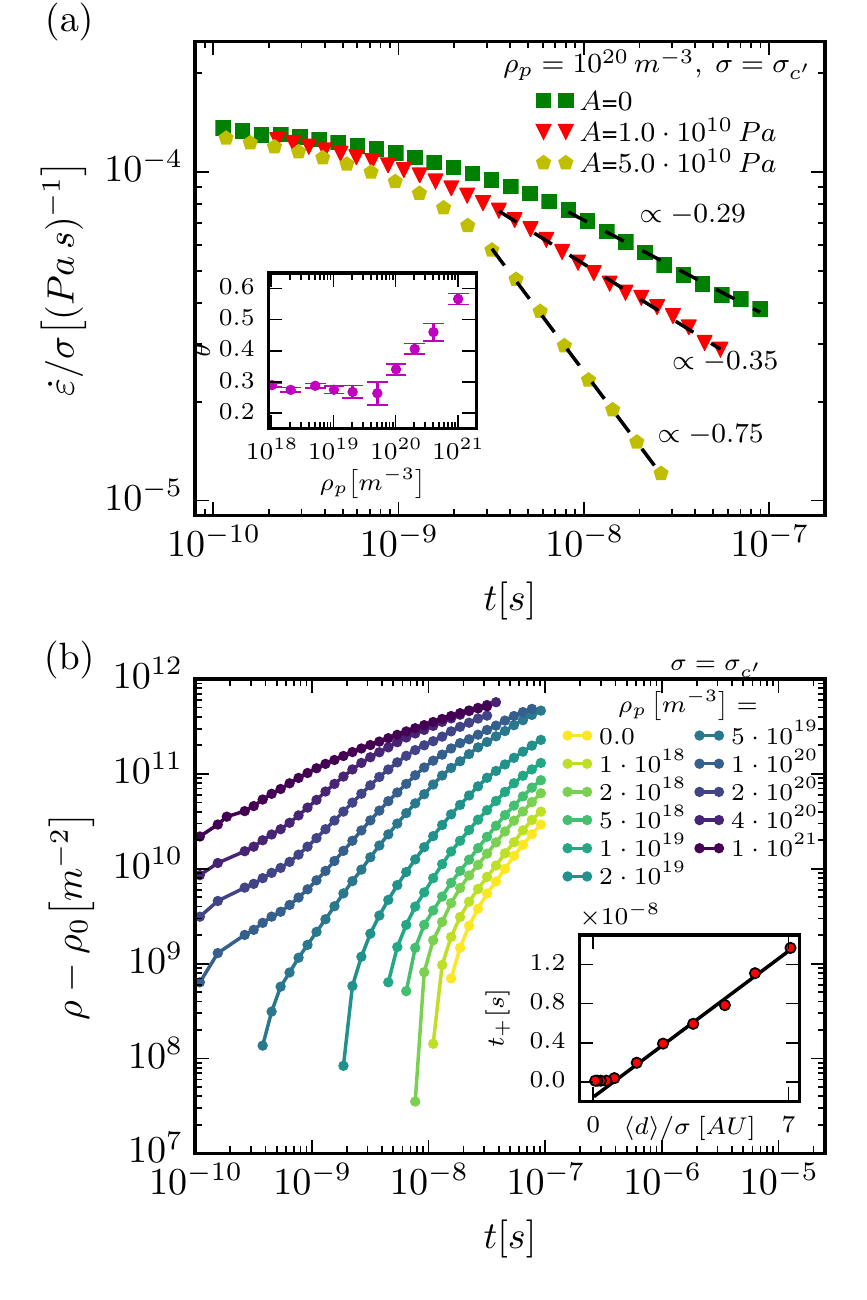}
  \caption{   (a) Different 
    $A$ produce different Andrade exponents $\theta$. Correspondingly the inset 
    shows $\theta (\rho_p)$. (b) The increase in dislocation density during the 
    creep with $\sigma=\sigma_{c'}$. The inset shows the existence of a correlation between the moment the density starts to increase, $t_{+}$, and 
    $\langle d \rangle$, the characteristic length scale of the dislocation and 
    precipitate configurations.}
  \label{fig2}
\end{figure}

The plastic flow in response to a constant $\sigma$ is characterized by a temporal Andrade power-law decay of the strain rate, $\dot{\varepsilon}(t) \propto t^{-\theta}$ \cite{miguel2002dislocation}. Examples of single system $\dot{\varepsilon}(t)$ of the illustrated systems in Fig. \ref{fig1}(a)-(c) and average $\dot{\varepsilon}(t)$
for different  $\sigma$ for the case with $A = 2.3 \cdot 10^{19}$
Pa m$^3$ and $\rho_p = 10^{20}$ m$^{-3}$ corresponding to Fig. \ref{fig1}(b) are shown in Figs. \ref{fig1}(d)-(e). These illustrate that the power-law 
decay of $\dot{\varepsilon}(t) \propto t^{-\theta}$ is obtained for a 
specific critical yield stress value $\sigma = \sigma_{c'}(A,\rho_p)$ only; 
For $\sigma < \sigma_{c'}$, $\dot{\varepsilon}(t)$ decays exponentially 
to zero, while for $\sigma > \sigma_{c'}$ the system appears to approach 
asymptotically a steady state with a finite
$\dot{\varepsilon} (\sigma)$. 
We repeat the simulations for a wide range of $A$ and $\rho_p$. The main panel of Fig. \ref{fig2}(a) shows three examples for different $A$-values of the critical relaxation of $\dot{\varepsilon}(t)$ at $\sigma = \sigma_{c'}(A)$. Interestingly, the exponent $\theta$ is found to increase with pinning strength $A$. The inset 
illustrates that upon increasing $\rho_p$ for a fixed $A$, the plasticity
exhibits two regimes: For pure and weakly 
disordered systems we find $\theta \approx 0.3$ independent 
of $\rho_p$, while for larger $\rho_p$ $\theta$ 
increases with $\rho_p$. The main panel of Fig. \ref{fig2}(b) shows how 
the dislocation density $\rho$ increases with time during the critical 
relaxation. A noticeable
increase from the initial value takes place at a
time scale $t_+$ which depends on $\rho_p$. The inset of Fig. \ref{fig2}(b)
shows that $t_+$ is linearly dependent on $\langle d \rangle/\sigma$, where
$\langle d \rangle = (d_p^{-3}+d_d^{-3})^{-1/3}$ ($d_p=\rho_p^{-1/3}$ and $d_d=\rho_0^{-1/2}$) is the characteristic defect-defect
distance, considering both dislocations and precipitates. Thus, $\rho$
starts increasing as soon as the dislocations get
pinned (by the nearest precipitate or another dislocation).
In the pinning-dominated regime $\rho_p \geq 10^{20}\,\text{m}^{-3}$, $\langle d \rangle$ is dominated by the precipitates, and increasing $\rho_p$ makes the relaxation faster, as evidenced by the $\rho_p$-dependent $\theta$. Interestingly, this is different from usual depinning transitions where the relaxation exponent is constant when varying microscopic detail \cite{laurson2011spatial}, which seems to be the case for the other critical yielding transition exponents below. These results may explain the wide range of experimentally observed $\theta$-values in for instance titanium alloys \cite{neeraj2000phenomenological}.

\begin{figure}[t!]
  \centering
    \includegraphics[width=0.38\textwidth]{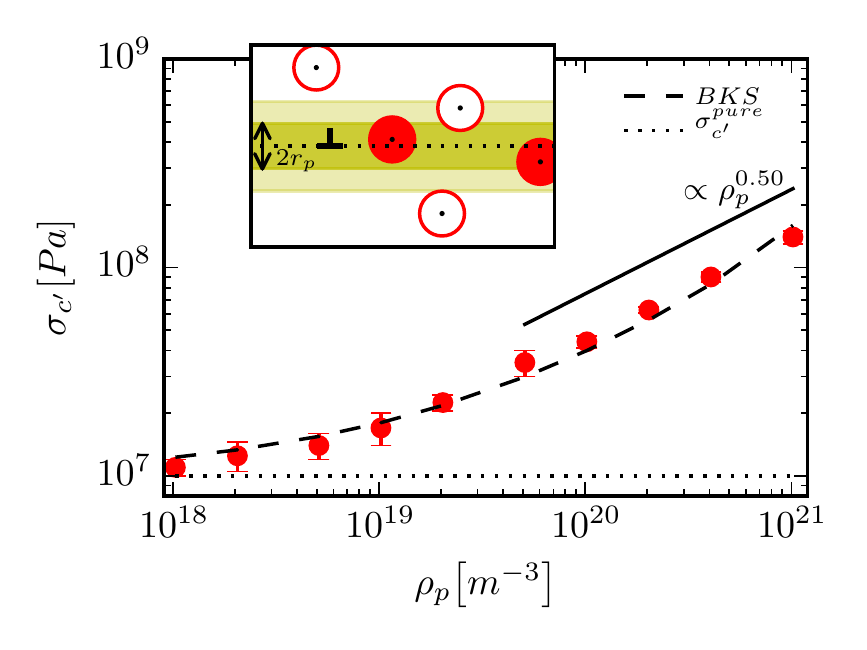}
  \caption{Dependence of $\sigma_{c'}$ on $\rho_p$ with $A=1.0 \cdot 10^{10}\,\text{Pa}$ with $\sigma_{c'}^{\text{pure}}$ (dotted line), the BKS equation (dashed line) and the asymptotic square root scaling with precipitate density (solid line).
  The inset: the mean precipitate distance in the glide plane:  the dashed line (main figure) is obtained by including the precipitates intersecting the glide plane  of the dislocation "$\perp$" (red circles with centers in the darker shaded region, compared to the white circles). %This is in contrast to previous suggestions to also count the precipitates in the lighter shaded region \cite{lehtinen2018effects}.
  }
  \label{fig3}
\end{figure}

For the yield stress we find an asymptotic power-law 
dependence $\sigma_{c'} \sim \rho_p^{0.50}$ for large $\rho_p$ (Fig. \ref{fig3}).
To comprehend this, we apply the Bacon-Kocks-Scattergood (BKS) relation connecting the obstacle hardening to the yield stress \cite{bacon1973effect,monnet2011orowan}.
The stress to overcome the obstacles is determined by the size and average distance of the randomly located obstacles along with the dislocation self-interactions. The BKS relation fits the simulation results very well [dashed line in Fig. \ref{fig3}] using the critical stress producing the power-law relaxation in creep $\sigma_{c'}$ as shown in Fig.\ \ref{fig1}(e). Also important is that one uses $ (2 r_p \rho_p)^{-1/2}$ as the mean inter-precipitate distance in the glide plane (Inset of Fig. \ref{fig3})  \cite{lehtinen2018effects} as a measure of the true interaction range. Thus 
\begin{equation}
\begin{aligned}{}
\sigma_{c'}(\rho_p) = \sigma_{c'}^{\text{pure}} + \frac{1}{2 \pi} \frac{G b}{\alpha \left[ (2 r_p \rho_p)^{-1/2} - 2 r_p \right]} \cdot \\ \left[ \ln{\frac{(2 r_p \rho_p)^{-1/2}}{r_\textrm{core}}} \right]^{-1/2} \left[ \ln{ \left( \frac{\overline{D}}{r_\textrm{core}} \right)}  + 0.7 \right]^{3/2},
\end{aligned}
\end{equation}
where $\sigma_{c'}^{\text{pure}}$ is the critical stress of a system without any precipitates, $\alpha$ is the Schmidt factor, $G$ the shear modulus, $b$ the magnitude of the Burgers vector, $r_\textrm{core}$ the dislocation radius and $\overline{D} = \frac{2 r_p (2 r_p \rho_p)^{-1/2}}{2 r_p + (2 r_p \rho_p)^{-1/2}}$.  The asymptotic square root dependence on $\rho_p$ is found [cf. Fig. \ref{fig3}(a)] in the depinning-dominated regime.

\section{Quasi-static loading simulations}

\begin{figure}[t!]
  \centering
    \includegraphics[width=0.38\textwidth]{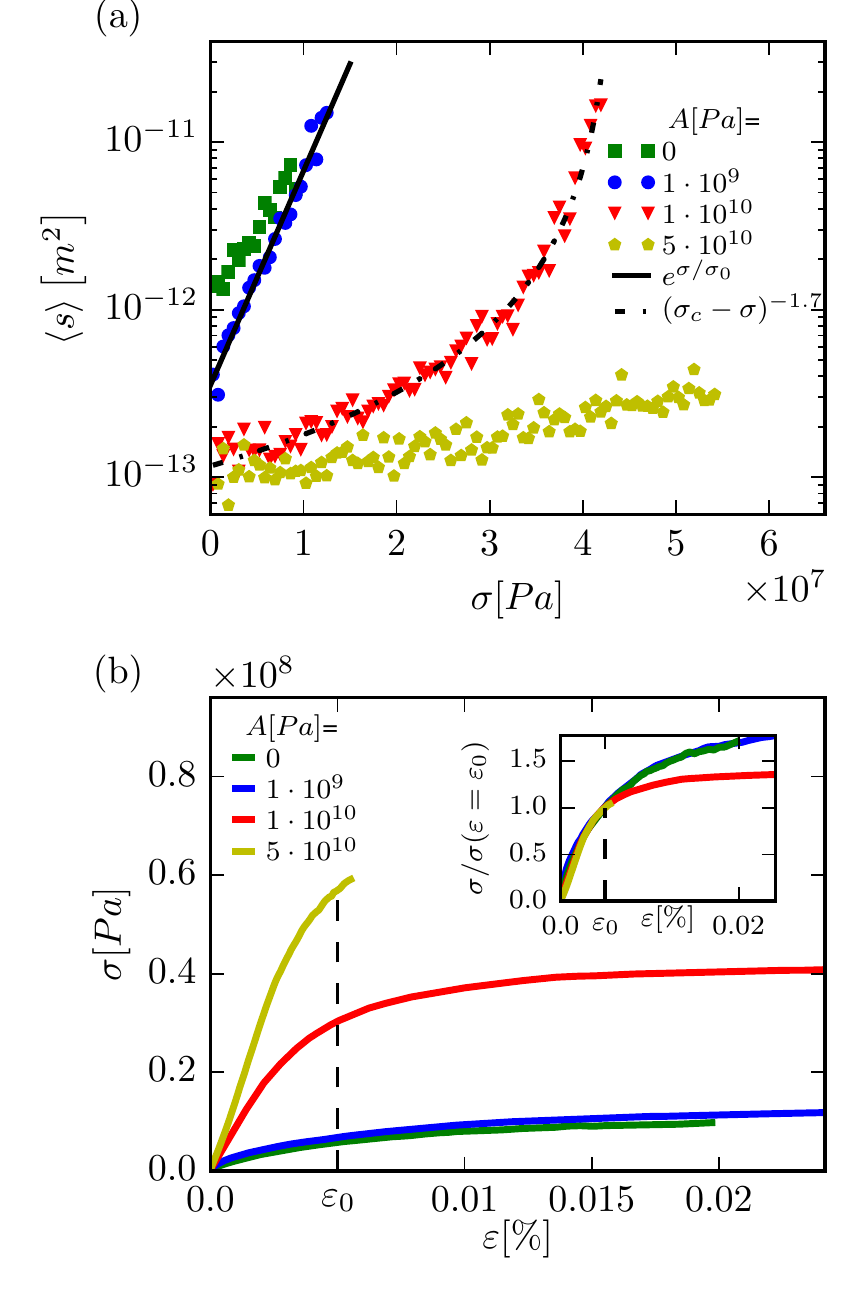}
  \caption{(a) The evolution of the average avalanche size as the external stress increases during the quasi-static loading for different values of $A$ and $\rho_p = 10^{20}\,\text{m}^{-3}$.
   The solid and dashed lines show the applied functions $\langle s \rangle \sim e^{\sigma/\sigma_0}$ and $\langle s \rangle \sim (\sigma_c- \sigma)^{-\gamma'}$, respectively. Here $\sigma_0=3.4 \cdot 10^6\,\text{Pa}$, $\sigma_c=4.4 \cdot 10^7\,\text{Pa}$ and $\gamma'=1.7$ were obtained by fitting.
   (b) Corresponding average stress-strain curves. The inset shows the same curves scaled by the stress value at $\varepsilon=\varepsilon_0=0.005\,\text{\%}$  (dashed line).  }
  \label{fig4}
\end{figure}

The transition from jamming to pinning is also seen in the dislocation avalanche statistics.
We fix $\rho_p$ to 10$^{20}$ m$^{-3}$ and vary $A$, and study the resulting 
avalanches $s$ seen in Fig.\ \ref{fig1}(f). 
Fig.\ \ref{fig4}(a) shows the mean size $\langle s \rangle$ as a 
function of $\sigma$ for different $A$-values. The systems with weakest 
disorder exhibit essentially the same exponential growth of $\langle s \rangle$ with 
$\sigma$ as pure 3D DDD systems \cite{lehtinen2016glassy}.
Weak precipitates are unable to pin the dislocations strongly enough to compete with dislocation jamming. They however make the system more susceptible to avalanches as evidenced by a decrease of the mean stress increment in between avalanches (Supplementary Fig.\ 2 \cite{SM}).

The evolution of $\langle s \rangle$ with $\sigma$ for intermediate strength disorder displays the typical behaviour of a depinning transition: 
First with small stress, $\langle s \rangle$ increases slowly, and appears 
to diverge when approaching $\sigma =\sigma_c \approx 4.4 \cdot 10^7\,\text{Pa}$
(Fig.\ \ref{fig4}(a)). It is harder to study the strongest precipitates due to numerical limitations in reaching high enough 
stresses/strains in that case to observe the expected divergence of 
$\langle s \rangle$. 

Fig.\ \ref{fig4}(b) depicts the average stress-strain curves with 
different disorder strengths.
As is evident, the increasing strength of 
precipitates hardens the system, so using an engineering criterion of stress at  $\varepsilon_0=0.005 \%$ we find an increasing yield stress with large enough $A$.
The inset of Fig.\ \ref{fig4}(b) displays the average stress-strain curves scaled by 
the above-defined yield stress $\sigma(\varepsilon=\varepsilon_0)$ which reveals that the shape of the curve changes drastically: with the weakest 
disorder or no disorder at all, the curves have the same shapes up to the maximum strain considered. Only for the intermediate disorder of 
$A=1.0 \cdot 10^{10}\,\text{Pa}$, with a depinning transition, the curve approaches a plateau 
value as the avalanche size diverges close to $\sigma_c$. 
For $A=5.0 \cdot 10^{10}\,\text{Pa}$, the shape of the curve is 
similar to the curve of intermediate disorder up to the reachable strain. The two kinds of average stress-strain
curves is what we would expect given that the avalanche statistics along the stress-strain curves are also different \cite{szabo2015plastic} as we discuss next.

\begin{figure*}[t!]
  \centering
    \includegraphics[width=0.8\textwidth]{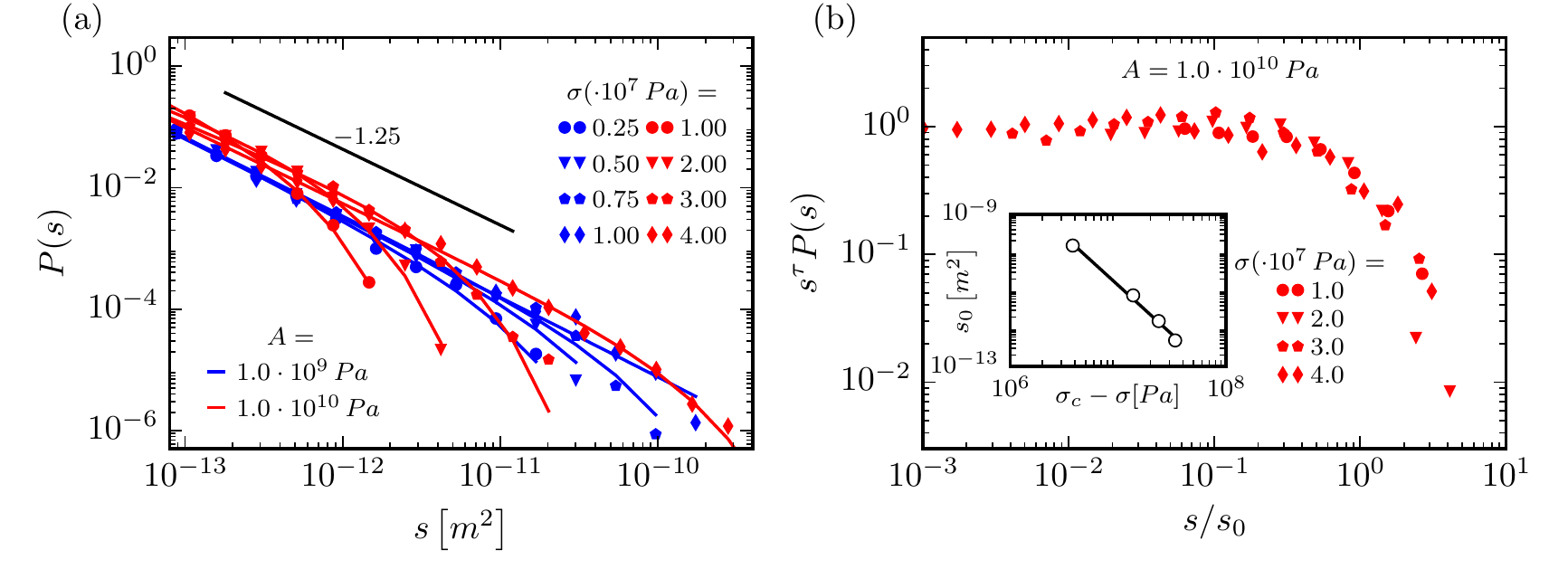}
  \caption{(a) The avalanche statistics from systems with precipitate parameters $\rho_p = 10^{20}\,\text{m}^{-3}$,  $A=1.0 \cdot 10^{9}\,\text{Pa}$ and $A=1.0 \cdot 10^{10}\,\text{Pa}$. The solid lines represent fits (see text). (b) shows the data collapse for $A=1.0 \cdot 10^{10}\,\text{Pa}$ and the inset illustrates the divergence of the distribution cutoff as $s_0 \sim (\sigma_c - \sigma)^{-1/\Sigma}$, where $\sigma_c=4.4\cdot 10^{7}\, \text{Pa}$ and $1/\Sigma=2.6$ were obtained by fitting. }
  \label{fig5}
\end{figure*}

Fig.\ \ref{fig5}(a) illustrates the stress-resolved size distributions of 
avalanches in weak and intermediate precipitate systems. The 
distributions of the avalanche sizes  are fitted with $P(s, \sigma) \propto s^{-\tau_s} e^{-s/s_0(\sigma)}$ where $s_0(\sigma)$ is a stress-dependent cutoff, and the best fit is chosen with the maximum likelihood method \cite{baro2012analysis}. 
The power law exponent $\tau_s$ does not vary much with $A$: 
$\tau_s=1.30\pm 0.02$ for $A=1.0 \cdot 10^{9}\,\text{Pa}$ and
$\tau_s=1.25\pm 0.02$ for $A=1.0 \cdot 10^{10}\,\text{Pa}$.
The behaviour of $s_0(\sigma)$
shows a clear dependence on $A$: With weak disorder
(or without disorder), there is only weak $\sigma$-dependence.
For intermediate $A$, 
Fig.\ \ref{fig5} shows a clear-cut divergence of the cutoff $s_0(\sigma)$
as the $\sigma_c$ is approached from below. 
This divergence is confirmed by the data collapse of Fig. \ref{fig5}(b) 
and its inset where $s_0 \propto (\sigma_c - \sigma)^{-1/\Sigma}$, with 
$1/\Sigma \approx 2.6$. Notably here the fitted value of $\sigma_c=4.4 \cdot 10^{7}\, \text{Pa}$ agrees well with the 
critical values of $\sigma_{c'}$ from creep in Fig.\ \ref{fig1}(e) and $\sigma_c$ in Fig.\ \ref{fig4}(a). The "engineering" yield stress from Fig.\ \ref{fig4}(b) is approximately $\sigma_c=3 \cdot 10^{7}\, \text{Pa}$ for this case. As a side note such a value is roughly four times larger 
than the $\sigma(\varepsilon=\varepsilon_0)$ in the limit of negligible precipitate strengthening.

For this depinning phase transition the exponent values $\tau_s \approx 1.25$ and $1/\Sigma 
\approx 2.6$ differ from those predicted by mean-field depinning (MFD; 3/2 and 2, 
respectively \cite{fisher1998collective}). The same applies to  %\cite{}
the stress-integrated size distribution ($\tau_{s,int}=1.39 \pm 0.01$ rather than 
2 for MFD, Supplementary Fig.\ 3 \cite{SM}), the exponent of the stress-resolved avalanche 
duration distribution ($\tau_{T}=1.30 \pm 0.05$, while MFD predicts 2, 
Supplementary Fig.\ 4 \cite{SM}), and the exponent characterizing the scaling of the 
average avalanche size with the avalanche duration, 
$\langle s \rangle \propto T^{\gamma}$, with $\gamma=1.76 \pm 0.01$ rather than 
2 for MFD (Supplementary Fig.\ 5 \cite{SM}). 
Interestingly, similar exponent values for $\tau_s$ and $\gamma$ have been recently found for amorphous plasticity \cite{budrikis2017universal}. It is also interesting to note that we recover the depinning of a single dislocation/elastic interface 
\cite{zapperi2001depinning, bako2008dislocation} by reducing 
the number of initial dislocations (Supplementary Fig.\ 6 \cite{SM}.) The important point here is that depinning of a single dislocation is different from the collective depinning of several dislocations we have focused on in the present study. \\

\section{Conclusions}
The yielding of precipitation-hardened FCC alloys demonstrate the role of collective phenomena in the nature of deformation bursts  
and in the averaged material response. The competition between the interactions among dislocations and those between dislocations and precipitates dictates the statistics of collective yielding  and controls the increase in yield strength by precipitates. Only when dislocation pinning due to disorder is the dominant mechanism %when the role of disorder is dominant mechanism 
the samples exhibit under loading a true critical point in the sense of a non-equilibrium phase transition.
A large increase in the flow stress follows when the disorder is relevant. The BKS relation for the yield stress might indeed be said to really work only for large precipitate densities when a yielding transition exists. We have here studied the case of fully coherent precipitates with isotropic interactions with the dislocations. However, the main point of competing mechanisms should persist so that similar disorder-dependent classes of critical behaviour are expected with other crystal orientations and structures and precipitate/metal systems. 

One should pay attention to how the concept of yield stress is used in this work. For quasi-static loading simulations
we use a typical way of defining it in DDD simulations, as the stress corresponding to a small amount of plastic strain.
However, we also use "true measures" of the yield stress which means that if there is a critical point, defined by the divergence of the avalanche size distribution, power-law relaxation of strain rate at constant stress, and continuous flow above the critical point,
then independent measurements of critical stress value should agree. The latter value is of course way higher and Figures \ref{fig3} and \ref{fig4} allow to compare such values.

The yielding or depinning critical point is characterized by non-trivial exponents different from those of MFD, as perhaps expected because of the anisotropic dislocation interactions \cite{budrikis2017universal,sparks2018nontrivial}. %rendering simple mean field descriptions inapplicable 
An open question is whether in this "strong disorder regime" one may find other universality classes, with different scaling properties for the avalanches. Finally, in micron-scale plasticity the strong dependence of the deformation fluctuations on details of the quenched pinning field becomes important for applications.

\begin{acknowledgments}
The authors acknowledge support from the Academy of Finland Center of Excellence program, 278367.
LL acknowledges the support of the Academy of Finland via the Academy Project COPLAST (project no. 322405), 
and AL that by the Academy Project SIRDAME (project no. 260053). 
HS acknowledges the support from Finnish Foundation for Technology Promotion. 
MA acknowledges support from the European Union Horizon 2020 research and innovation
programme under grant agreement No 857470 and from European Regional Development Fund via Foundation for Polish Science International Research Agenda PLUS programme grant No MAB PLUS/2018/8.
The authors acknowledge the computational resources provided by the Aalto University 
School of Science ``Science-IT'' project, as well as those provided by CSC (Finland).%\\
\end{acknowledgments}

 \bibliographystyle{apsrev}
 \bibliography{refs}

\end{document}